\documentclass[fleqn,usenatbib]{mnras}

\usepackage{newtxtext,newtxmath}
\usepackage[T1]{fontenc}

\DeclareRobustCommand{\VAN}[3]{#2}
\let\VANthebibliography\thebibliography
\def\thebibliography{\DeclareRobustCommand{\VAN}[3]{##3}\VANthebibliography}

\usepackage{array,multirow,graphicx} % Including figure files
\usepackage{float} 
\usepackage{amsmath}  % Advanced maths commands
\usepackage{physics}
\usepackage{tabularx}
\usepackage[dvipsnames]{xcolor}
\usepackage{tikz}
\usepackage{orcidlink}
\usepackage{hyperref}
\newcommand\REPLY[1]{\textcolor{black}{#1}}
\title[Lens search in two JWST fields]{Galaxy-scale lens search in the PEARLS NEP TDF and CEERS JWST fields}

\author[G. Ferrami et al.]
{
Giovanni Ferrami$^{1, 2}$\thanks{E-mail: gferrami@student.unimelb.edu.au }\orcidlink{0000-0003-1268-5230}, 
Nathan J. Adams$^{3}$\orcidlink{0000-0003-4875-6272}, 
Lewi Westcott$^{3}$\orcidlink{0009-0008-8642-5275},
Thomas Harvey$^{3}$\orcidlink{0000-0002-4130-636X},
Rolf A. Jansen$^{4}$\orcidlink{0000-0003-1268-5230},
\newauthor Jose M. Diego$^{5}$\orcidlink{0000-0001-9065-3926},
Vince Estrada-Carpenter$^{4}$\orcidlink{0000-0001-8489-2349}, 
Rogier A. Windhorst$^{4, 6}$\orcidlink{0000-0001-8156-6281}, 
Christopher J. Conselice$^{3}$\orcidlink{0000-0003-1949-7638},
\newauthor Anton M. Koekemoer$^{7}$\orcidlink{0000-0002-6610-2048},
Jordan C. J. D’Silva$^{8}$\orcidlink{0000-0002-9816-1931}, 
Christopher Willmer$^{9}$\orcidlink{0000-0001-9262-9997}, 
J. Stuart B. Wyithe$^{10, 2}$\orcidlink{0000-0001-8156-6281},
\newauthor Michael J. Rutkowski$^{11}$\orcidlink{0000-0001-7016-5220}, 
Seth H. Cohen$^{4}$\orcidlink{0000-0003-3329-1337},
Brenda L. Frye$^{9}$\orcidlink{0000-0003-1625-8009},
and Norman A. Grogin$^{7}$\orcidlink{0000-0001-9440-8872}
\\
\\
$^{1}$School of Physics, University of Melbourne, Parkville, VIC 3010, Australia\\
$^{2}$ARC Centre of Excellence for All-Sky Astrophysics in 3 Dimensions (ASTRO 3D), Canberra, ACT 2611\\
$^{3}$Jodrell Bank Centre for Astrophysics, Alan Turing Building, University of Manchester, Oxford Road, Manchester M13 9PL, UK\\
$^{4}$School of Earth \& Space Exploration, Arizona State University, Tempe, AZ 85287-1404, USA\\
$^{5}$Instituto de Fisica de Cantabria (CSIC-UC). Avenida. Los Castros s/n. 39005 Santander, Spain\\
$^{6}$Department of Physics, Arizona State University, Tempe, AZ 85287-1504, USA\\
$^{7}$Space Telescope Science Institute, 3700 San Martin Drive, Baltimore, MD 21218, USA\\
$^{8}$International Centre for Radio Astronomy Research (ICRAR), The University of Western Australia, M468, 35 Stirling Highway, Crawley, WA 6009, Australia\\
$^{9}$Steward Observatory, University of Arizona, 933 North Cherry Avenue, Tucson AZ, 85721, USA\\
$^{10}$Research School of Astronomy and Astrophysics, Australian National University, Canberra, ACT 2611, Australia\\
$^{11}$Minnesota State University-Mankato, Dept. of Physics \& Astronomy, Trafton Science Center North 141, Mankato, MN, 56001 USA
}

\date{Accepted XXX. Received YYY; in original form ZZZ}

\pubyear{\the\year{}}

\begin{document}
\label{firstpage}
\pagerange{\pageref{firstpage}--\pageref{lastpage}}
\maketitle

\begin{abstract}
We present four galaxy scale lenses discovered in two JWST blank-fields: the $\sim$ 54 arcmin$^2$ of the {PEARLS} North-Ecliptic-Pole Time-Domain Field (NEP TDF) and in the $\sim 90$ arcmin$^2$ of {CEERS}.
We perform the search by visual inspection of NIRCam photometric data, obtaining an initial list of 16 lens candidates.
We down-select this list to \REPLY{5} high-confidence lens candidates, based on lens modelling of the image configuration and photometric redshift measurements for both the source and the deflector.
We compare our results to samples of lenses obtained in ground-based and space-based lens searches and theoretical expectations.
We expect that JWST observations of field galaxies will yield approximately 1 galaxy scale lens every three to \REPLY{four} NIRCam pointings of comparable depth to these observations ($\sim$ 9 arcmin$^2$ each).
This shows that JWST, compared to other lens searches, can yield an extremely high number of secure lenses per unit area, with redshift and size distributions complementary to lens samples obtained from ground-based and wide-area surveys.
We estimate that a single JWST pure-parallel survey of comparable depth could yield $\sim \REPLY{80}$ galaxy scale lenses, with a third of them having $z_{\rm{lens}}>1$ and $z_{\rm{source}}>3$.
\end{abstract}


%%%%%%%%%%%%%%%%%%%%%%%%%%%%%%%%%%%%%%%%%%%%%%%%%%

%%%%%%%%%%%%%%%%% BODY OF PAPER %%%%%%%%%%%%%%%%%%

\section{Introduction}
Strong gravitational lensing has proved to be a flexible tool to probe astrophysics and cosmology simultaneously, as it depends on the total projected mass distribution of the object acting as a lens and the cosmological distances of the lens and the background source.
In particular, galaxy scale lenses can constrain the inner structure of galaxies within $\sim 1$ arcsecond (e.g., \citealt{Treu_Koopmans_2002}, \citealt{Sonnenfeld_SL2S_2013}, \citealt{Dinos_1}), constrain the redshift evolution of the velocity dispersion function (\citealt{Kochanek_1992,Mitchell_2005_VDF_from_LF,Capelo_Natarajan_2007,Ferrami_VDF}), bias the observed galaxy luminosity function (e.g., \citealt{Barone_Nugent_2015}, \citealt{Mason_2015},  \citealt{Ferrami_Lensing_bright_end}), measure the values of cosmological parameters from images time-delays (e.g., \citealt{Refsdal_1964}, \citealt{Shajib_lensing_H0}) or multiple plane lenses (e.g. \citealt{Collett_Auger_DoubleSrc}).
To date, we know around $\mathcal{O}(10^3)$ galaxy-scale lenses (many of which are only candidates that require follow-up confirmation), and therefore analyses based on samples of strong lenses have often been limited by small-number statistics.

To extract a sample of galaxy scale lenses from the photometric data, a range of lens search methods have been tested: algorithms for feature detection (e.g., \citealt{ARCFINDER_More}, \citealt{RINGFINDER_ALGORITHM})
lens model fitting algorithms (e.g. \citealt{Marshall_model_detection});
or a combination of the two (e.g., \citealt{Sonnenfeld_SuGOHI_I}); 
visual search by researchers (\citealt{Hogg_lens_in_HST,Moustakas_visual_inspection,Faure_2008,COWLS_I}); citizen science (\citealt{Marshall_citizien_science}; \citealt{More_citizien_science};  \citealt{Sonnenfeld_citizien_science}).
In addition, various machine learning architectures have been employed so far in lens searches  Convolutional Neural Networks (CNNs, see for example \citealt{Jacobs_CNN_DES}, \citealt{EUCLID_CNN_lensfinder}), Support Vector Machines (SVM, e.g., \citealt{Support_Vector_Machines}) and self-attention encoding (e.g., \citealt{Self_attention_ML}).

The degree of completeness and purity of a particular lens search method when applied to real data is hard to reliably estimate, and visual inspection can outperform other forms of automated search, such as neural networks (e.g., \citealt{Holloway_ensemble_classifier}, which also shows that combining visual inspection with CNNs can improve the final sample).

By the end of this decade, wide-field photometric surveys are expected to yield $\mathcal{O}(10^5)$ strong lenses, two orders of magnitude more than the current sample (\citealt{Collett_2015}, \citealt{Holloway_2023}, \citealt{Ferrami_lensstat}).
These surveys include the Euclid Wide survey (\citealt{Euclid_Wide_Survey}), Vera Rubin Observatory LSST (\citealt{Vera_Rubin_LSST}), and Roman Space Telescope High Latitude Wide Area Survey (\citealt{Roman_WFIRST}).
Compared to these large surveys, a typical field observed by James Webb Space Telescope (JWST) photometry is smaller, but much deeper in flux, with much better spatial resolution, and with several near-infrared (NIR) filters that can see faint red sources behind bright deflectors.
These characteristics allow JWST to probe a different regime of lenses, both in redshift and angular size.

In this article, we report on a visual search of galaxy scale lenses in two JWST fields of comparable size, PEARLS North Ecliptic Pole Time Domain Field (NEP TDF) 54 arcmin$^2$ and CEERS $\sim 90$ arcmin$^2$, both with ancillary Hubble Space Telescope (HST) optical photometry to enable accurate photometric redshift estimation.
We find \REPLY{5 high-confidence} candidates, at lower angular sizes and higher redshifts than lens samples identified by ground-based and optical surveys, in agreement with theoretical expectations, and with recent findings of the \REPLY{COSMOS-Web Lens Survey (COWLS)} lens search (\citealt{COWLS_I, COWLS_II, COWLS_III}).

This article is structured as follows. 
In Section \ref{Section:Data} we describe the reduction of the photometry of the two fields.
In Section \ref{Section:LensSample} we introduce the strategy for visual inspection, SED fitting and  lens modelling.
In Section \ref{Section:Results} we present the results of the lens search and discuss how the resulting distributions compare with previous surveys and theoretical expectations.
Finally, in Section \ref{Section:Conclusions} we present our conclusions.

Throughout this paper, we adopt $H_0 = 70$ km s$^{-1}$ Mpc$^{-1}$, $\Omega_0 = 0.3$,  $\Omega_\Lambda= 0.7$.

\setcounter{footnote}{0}

\section{Data}\label{Section:Data}
In this paper we used publicly avaiable HST and JWST data drawn from the Prime Extragalactic Areas for Reionization and Lensing Science (PEARLS, PIs: R. Windhorst \& H. Hammel, PIDs: 1176 \& 2738, \citealt{windhorst2023}) North Ecliptic Pole Time-Domain Field (NEP TDF; \citealt{jansenwindhorst2018}) and the Cosmic Evolution Early Release Science (CEERS, PID: 1345, PI: S. Finkelstein, \citealt{CEERS_Finkelstein, CEERS_Bagley}) programs.
These two fields were chosen because both: 
\begin{itemize}
\item are blank-fields (i.e., not centred on a cluster), allowing us to study the galaxy-scale strong lensing statistics,
\item have footprints smaller than $\sim100$ arcmin$^2$, with deeper photometry to detect the faint light of lensed background sources over a large range of redshifts, 
\item contain at least 9 photometric filters from JWST/NIRCam and HST/ACS combined, to enable robust SED fitting.
\end{itemize}

\subsection{PEARLS NEP Time-Domain Field}
This field is located within JWSTs continuous viewing zone, meaning it is accessible to JWST and HST at any time of the year. The PEARLS data consists of 8 partially overlapping NIRCam pointings that are grouped into 4 pairs. Each pair of pointings is taken 3 months apart, resulting in a cross/windmill shaped geometry for the field. NIRCam data were reduced as part of the EPOCHS project, using JWST pipeline version 1.8.2, calibration pmap1084 and a pixel scale of 0.03as/pix \citep{Adams2024,Conselice2024}. In addition to standard calibrations, we employ the use of a 1/f correction developed by Chris Willott \footnote{\url{https://github.com/chriswillott/jwst/tree/master}} and subtract scaled templates of `Wisp' artefacts in the F150W and F200W images. There are a total of 8 NIRCam bands spanning $\sim60$ square arcmins, including F090W, F115W, F150W, F200W, F277W, F356W, F410M, F444W. We calculate point-source depths ranging from 28.5-29.3 within 0.32as diameter apertures ($\sim80\%$ enclosed PSF flux) from the random scattering of empty apertures across the final mosaic.
Part of this field is also covered by NIRISS slitless spectroscopy (Estrada-Cartpenter et al. [in prep.]).
While in this work we focus on the photometric redshift obtained from the SED fitting over the available filters, the 
NIRISS spectra of one of the lenses found in our sample (PEARLS J172238.9+655143) is analysed in \cite{NathanAdams_EinsteinRingInPEARLS}. 

To aid photometric analysis, we include HST imaging in the F606W filter from the TREASUREHUNT programme \citep[][Jansen et al. in prep., GO 15278, PI: R. Jansen and GO 16252/16793, PIs: R. Jansen \& N. Grogin]{Obrien2024}, pixel matched to the JWST imaging. 
Data spans from Oct 2017 to Oct 2022 and provides F275W, F435W and F606W ACS imaing over 194 square arcminutes. The F606W mosaic has a comparable depth to the JWST data, with a limiting $5\sigma$ magnitude of 28.5 in the same aperture size of 0.32as, whilst the bluer bands are upwards of 1 magnitude shallower. For this reason, we only employ the use of the F606W mosaic in this study. Details of HST data reduction, as well as the general field/survey geometry, are provided in \citet{Obrien2024}.

\subsection{CEERS}
This study also makes use of both primary observing runs (July 2022 \& December 2022) of the CEERS survey. 
This consists of 10 NIRCam pointings mosaiced over the Extended Groth Strip (EGS: \citealt{ExtendedGrothStrip}) with 7
photometric bands (F115W, F150W, F200W, F277W, F356W, F410M and F444W). We use the mosaics produced using the same pipeline as our NEP-TDF images described above. 
This field provides the a contigous area of 64.15 square arcminutes after masking and it reaches depths between apparent magnitudes of 28.6 and 29.3. The field does lack the F090W band and so we include HST CANDELS imaging of the F606W and F814W filters that was reduced by the
CEERS team (their HDR1) following \cite{CANDELS_Koekemoer} in order to cover this bluer wavelength range.

\section{Identifying the lens sample}\label{Section:LensSample}
\begin{figure*}
\includegraphics[width=\linewidth]{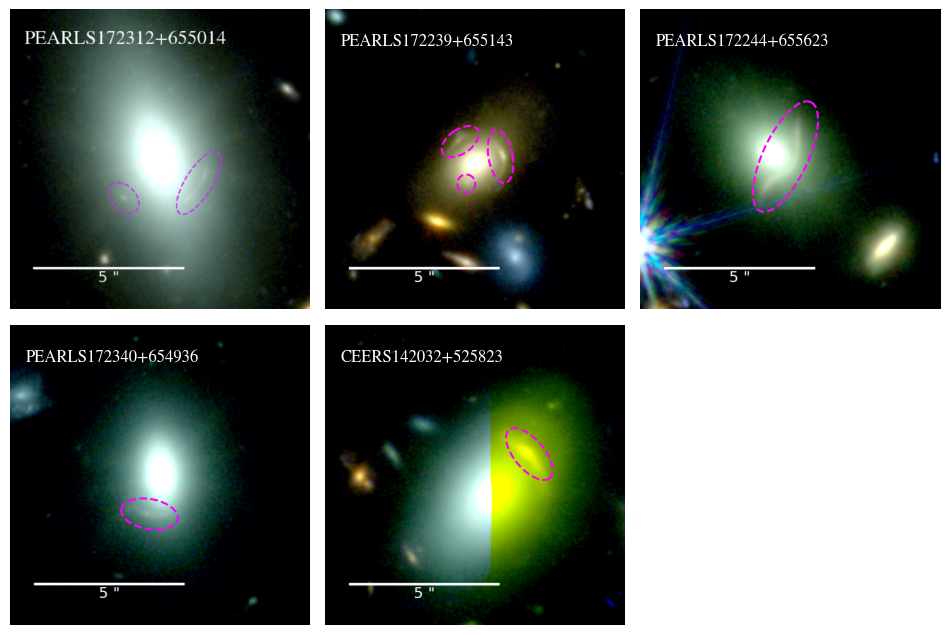}
    \caption{Lens candidates in {PEARLS} NEP field and CEERS fields. The cutouts are 10"$\times$10", and the magenta contours highlight the potential lensing features noticed in the visual inspection process.
    We note that the main arc in CEERS J142031.8+525822 lies in the region where there is no blue NIRCam data.}
    \label{fig:lens_candidates}
\end{figure*}

The process we followed to identify the \REPLY{5} lenses presented in this work started with a visual inspection of the PEARLS NEP-TDF and CEERS fields, which yielded an initial set of 16 candidate lenses (see Appendix and Figure \ref{fig:lens_candidates_initial}).
Classifying a candidate as a genuine lens requires that the geometry of the system is compatible with a lensing system (i.e. the deflector and the source lie at different redshifts, to be confirmed spectroscopically), and a model for the total projected mass and source light that could reproduce the observed strong lensing features.
One of our candidate systems, PEARLS J172238.9+655143, is located within the NIRISS F200W field of view. A detailed analysis of this system, including its spectra, can be found in \cite{NathanAdams_EinsteinRingInPEARLS}. In this paper, we relied on photometric redshifts of both deflector and source of each candidate and we fitted the available photometry to a parametric lens model.
These steps lead us to final list of \REPLY{5} high-confidence lens candidates, shown in Figure \ref{fig:lens_candidates}.
The visual inspection step, the SED fitting procedure (necessary to obtain the redshift estimates), and lens fitting process are described in the following subsections.

\subsection{Visual Inspection}
Since the total area obtained combining the PEARLS NEP-TDF and CEERS fields is around $144$ arcmin$^2$, we were able to conduct a visual inspection of the whole field, without restricting the search around massive galaxies (i.e., without introducing luminosity cuts to the deflector population).
Visual inspection was conducted by GF, RAJ, and JMD in the NEP-TDF, and by GF in CEERS.
The inspectors looked at the coloured images of all JWST pointings constructed by mapping the light intensity observed in the F444W, F277W, and F150W bands for the red, green, and blue channels, respectively.
Each field was inspected twice to obtain an initial list of 16 candidates that showed some morphological feature that looked possibly due to strong lensing. 
Half of these candidates lie in the NEP-TDF field and half in CEERS.

\subsection{SED modelling and photometric redshift}
We model the foreground elliptical of each initial candidate by providing PSF homogenised images to the {\tt galfit} code \citep{Peng2010}. PSF homogenisation was conducted with the use of {\tt pypher} \citep{boucaud2016} and model PSFs are generated following an empirical approach for both the JWST and HST imaging \citep{Skelton2014,Whitaker2019,Weaver2024}. 
We use the underlying image weight maps to identify and cut out a large region around our source which uses the same combination of mosaiced images to produce our empirical PSF model. 
To account for this, and following the methods and work in \citet{Westcott2024}, single and double Sérsic profiles are fit to the elliptical galaxy with surrounding sources masked in each photometric band. Using this method we find that a single Sérsic profile fits the source well. 
To obtain an initial photometric redshift for the foreground galaxy, we provide the SED fitting code {\tt EAZY}\citep{Brammer2008} with aperture photometry extracted from the light profile fit using the base set of FSPS models. 
For the higher-z background galaxy, we trial the use of both default FSPS models in {\tt EAZY} \citep{Conroy2009,Conroy2010} in addition to templated featuring bluer SEDs and stronger emission lines derived in the work by \citep{Larson2022} and obtain similar redshifts.
For the process of obtaining an initial photo-z estimate, photometric errors are fixed to 10\%. For stellar masses, fluxes extracted within circular apertures of radius $\theta_E$ are run through the SED fitting tool {\tt BAGPIPES} \citep{Carnall2018} using a fiducial Kroupa IMF and the same set up as \citep{Harvey2025}. To summarise here; we use LogNormal star formation histories and fix the redshift to the spec-z if available or use a prior following a Gaussian approximation of the {\tt EAZY} $P(z)$ otherwise. \citet{Calzetti2000} dust extinction up to 3.5 mags is allowed and broad, uniform priors on other parameters such as metallicity.

\subsection{Lens modelling}
\begin{figure*}
\includegraphics[width=1\linewidth]{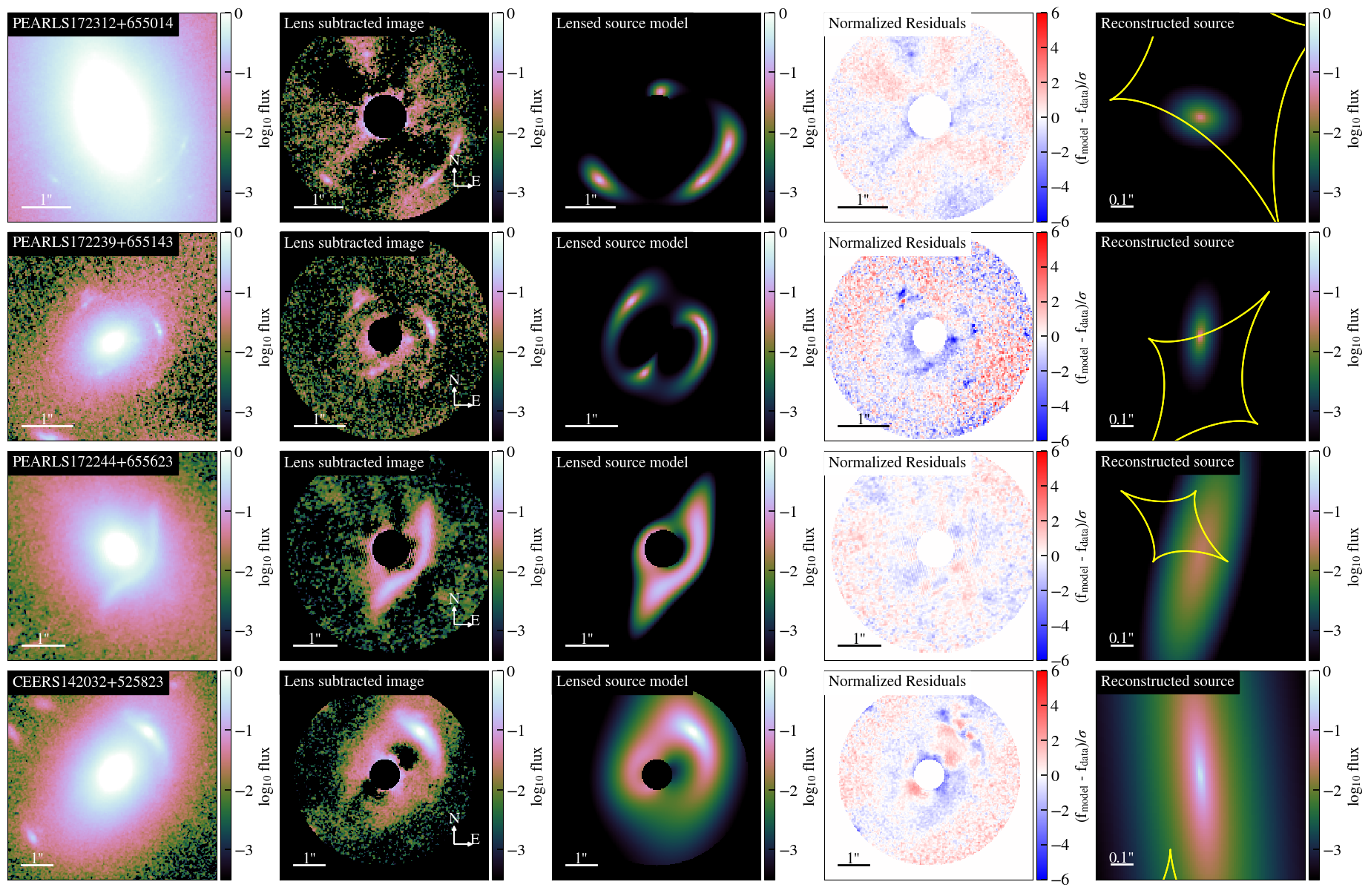}
\caption{The four lenses fit with a SIE mass distribution plus external shear. 
From left to right, the panels show the single-band image data (F444W) for each lens, \REPLY{the data where a S\'ersic model subtracts the foreground lens emission, the image–plane lensed source model,} the residual difference between the image, and the reconstructed source plane with the tangential caustic highlighted in yellow.}
\label{fig:all_models}
\end{figure*}

To confirm that the observed image configuration is compatible with strong lensing, we model each system with a simple parametric model. 
We model the deflector as a Singular Isothermal Ellipsoid (SIE) mass distribution with external shear, and the deflector and background source light as an elliptical Sérsic profile \citep{Sersic_profile}.
The total mass distribution within the Einstein radius of early-type galaxies (ETGs) is well approximated by an isothermal distribution, such as the SIE (e.g, \citealt{Gavazzi_SLACS_2007}, \citealt{Koopmans_bulge_halo_conspiracy}, \citealt{Lapi_2012}, \citealt{SLACS_debiased}). 
The external shear component accounts for additional shear introduced by line-of-sight perturbers and compensate for the simplicity in the angular structure of the main deflector model (e.g. \citealt{EtheringtonExtShear}).
To perform the parametric lens fitting we use the {\tt lenstronomy} package \citep{Birrer2018,Birrer2021}.
In the fitting process, we introduce an appropriately sized annular mask to avoid large residuals in the centre of the lens light profile (within $\approx 0.3$", where the Sérsic profile is less adequate in reproducing the observations), and block out the light from the line-of-sight galaxies in the outer region.
We sampled from the posterior probability distribution function (PDF) of the model parameters with the Markov chain Monte Carlo (MCMC) affine invariant method \citep{Goodman_Weare} using {\tt emcee} \citep{emcee}. To achieve faster convergence in the MCMC sampling, we choose the initial points of the Markov-chain from the best-fitting parameters obtained with the particle swarm optimization (PSO) method \citep{ParticleSwarmOptimisation}.
The SED and lens modelling yielded 4 high-confidence lens candidates, \REPLY{with their} lens models shown in Figure \ref{fig:all_models}.

\subsection{Alternative lens-light subtraction and modelling}
\begin{figure*}
\includegraphics[width=1\linewidth]{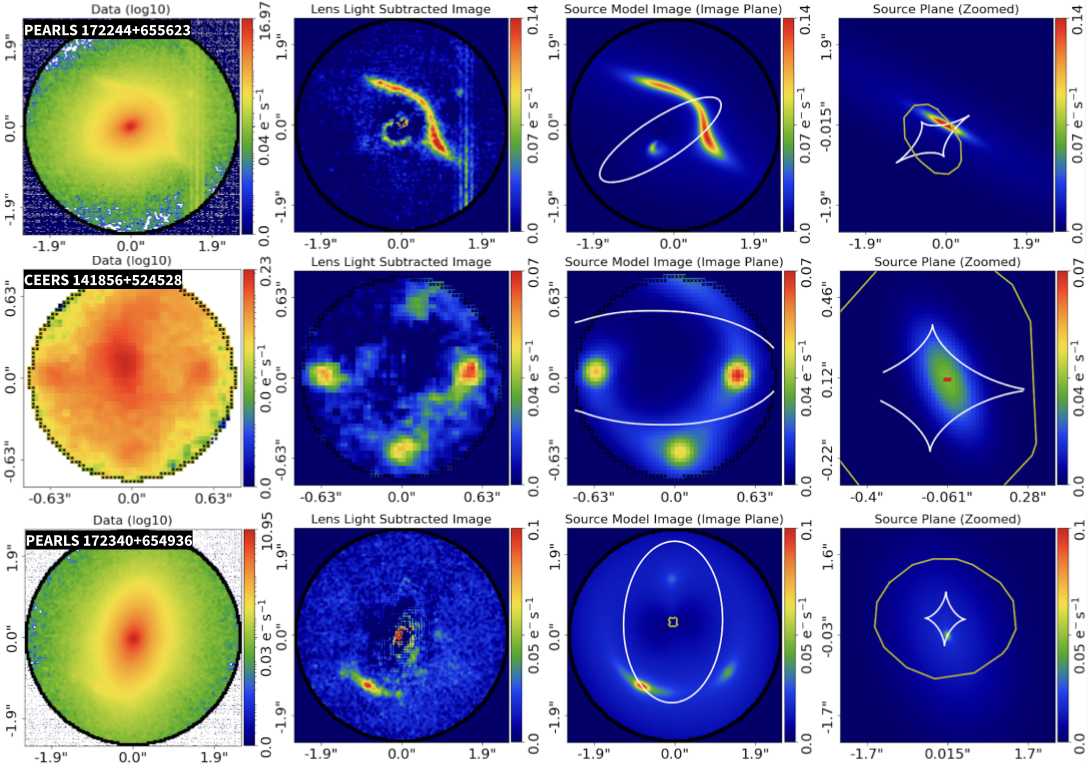}
\caption{
\REPLY{Example of three systems modelled using the pipeline similar to the one employed in COWLS (\citealt{COWLS_I}). From left to right, the panels show the single-band image data (F444W), a foreground-subtracted image using a Multi-Gaussian Expansion (note that the centre of the cutout is not masked), the model lensed source in the image plane, and source-plane reconstructions.}
}
\label{fig:COWLS_like_analysis}
\end{figure*}

\REPLY{
While applying an annular mask to a lensing system often enables the lens light to be modelled with a small number of parameters using an elliptical S\'ersic profile, the central mask can obscure secondary images that commonly form near the centre of the deflector. Moreover, at the resolution of space-based observatories such as HST or JWST, a S\'ersic profile may be insufficient to capture the complex morphologies exhibited by some deflectors. To mitigate this limitation, we re-model our sample using an approach similar to that adopted in the COWLS lens search. (\citealt{COWLS_I}).
We decompose the lens light in a set of 2D elliptical Gaussian profiles with a Multi-Gaussian Expansion (\citealt{MGE_Cappellari}, hereafter MGE). We use the implementation described in \cite{MGE_Pyautolens} included in the {\tt pyautolens} package (\citealt{Pyautolens_I, Pyautolens_II, Pyautolens_III}).
Following \cite{COWLS_I}, we group the MGE light model in three sets, where the components of each set share the same centre, position angle and axis ratio.
We restrict two sets of 30 Gaussians to have their standard deviation $\sigma_{\text{MGE}}$ logarithmically spaced between one-fifth of the image pixel scale and the circular radius of the outer mask (which is kept the same as the previous modelling step), and a third set of 10 Gaussians with $\sigma_{\text{MGE}}$ spaced between $0.01$ arcsec and twice the pixel scale.
For each lens candidate, the lens-subtracted cutouts are then used to fit a lens model consisting of a SIE mass distribution with external shear using {\tt pyautolens}. The lensed sources are modelled as an MGE composed of a set of 30 Gaussians with $\sigma_{\text{MGE}}$ spaced between $0.001$ and $1$ arcsec. This modelling strategy yields one additional high-confidence lens, PEARLS J172339.6+654936, and an example of three systems modelled this way is shown in Figure \ref{fig:COWLS_like_analysis}.
This method also yields a qualitatively good model for CEERS J141855.6+524527, which is not included in our high-confidence sample due to images and background being at very similar redshifts based on our photo-z estimates (see also Table \ref{tab:Photo_z_all_candidates}).
}

\section{Results of the lens search}\label{Section:Results}

The position and redshift of the final sample of \REPLY{5} lenses are summarised in Table \ref{tab::Discovered_lenses}, and the parameters of the best fitting lens model and stellar mass are listed in Table \ref{tab::Lens_models_parameters}.
The lens redshift and source redshift ranges are $0.38 < z_{\rm{lens}} < 1.25$ and $1 < z_{\rm{source}} < 3.6$, respectively. Their Einstein radii range between 0.7"$<\theta_E<$1.3", and the velocity dispersion entering the SIE model ranges between $200 < \sigma/(\rm{km/s}) < 250$, consistent with the predictions of the fiducial model in \cite{Ferrami_lensstat}.
The NEP TDF field lens PEARLS J172238.9+655143 has a photo-z for the background source of $\sim 3.6$, amongst the highest in known galaxy-scale deflectors. A more detailed analysis of this system can be found in \cite{NathanAdams_EinsteinRingInPEARLS}.

After having identified the \REPLY{5} high confidence lenses in our sample, we cross checked previous lens searches and found that CEERS J142031.8+525822 was also found in the SL2S lens search \citep{Sonnenfeld_SL2S_III_2013, Sonnenfeld_SL2S_2013} (identified there as SL2SJ142031+525822 and classified as grade B, potential lens).
We choose to use the spectroscopic redshifts for deflector and source presented in SL2S in Table \ref{tab::Discovered_lenses}, also because this lens falls partially in the gap of the bluest NIRCam filter available.

The projected stellar and total mass estimates, obtained from {\tt BAGPIPES} and the best fitting lens model respectively, are presented in Table \ref{tab::Mass_within_REinstein}. The range of stellar-to-total mass ratio in our sample is consistent with the dark matter fraction found in other sample of lenses (e.g. in SL2S, \citealt{Sonnenfeld_SL2S_darkmatter}). We do not explore the contribution of different choices of IMF on the inferred mass ratios as our sample is limited to photometric redshifts, which could bias the total projected mass (e.g \citealt{ShuntovCosmosRing}).

\begin{figure*}
\includegraphics[width=\linewidth]{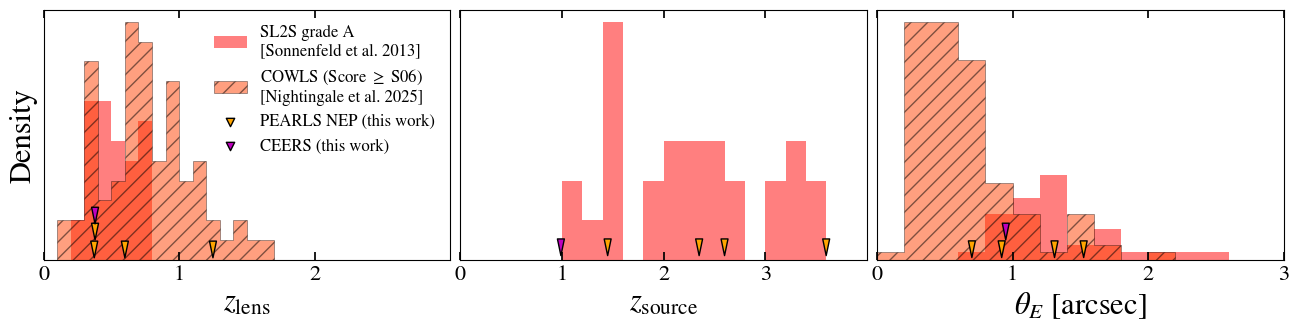}
\caption{Properties of our lens sample (arrows) compared with the lenses in SL2S (red, \citealt{Sonnenfeld_SL2S_2013}) and COWLS (orange dashed, \citealt{COWLS_I}). 
The left and middle panels show the deflector and source redshift distributions, respectively. The right panel shows the Einstein radius distribution.}
\label{fig:comp_distributions}
\end{figure*}

\begin{figure*}
\includegraphics[width=\linewidth]{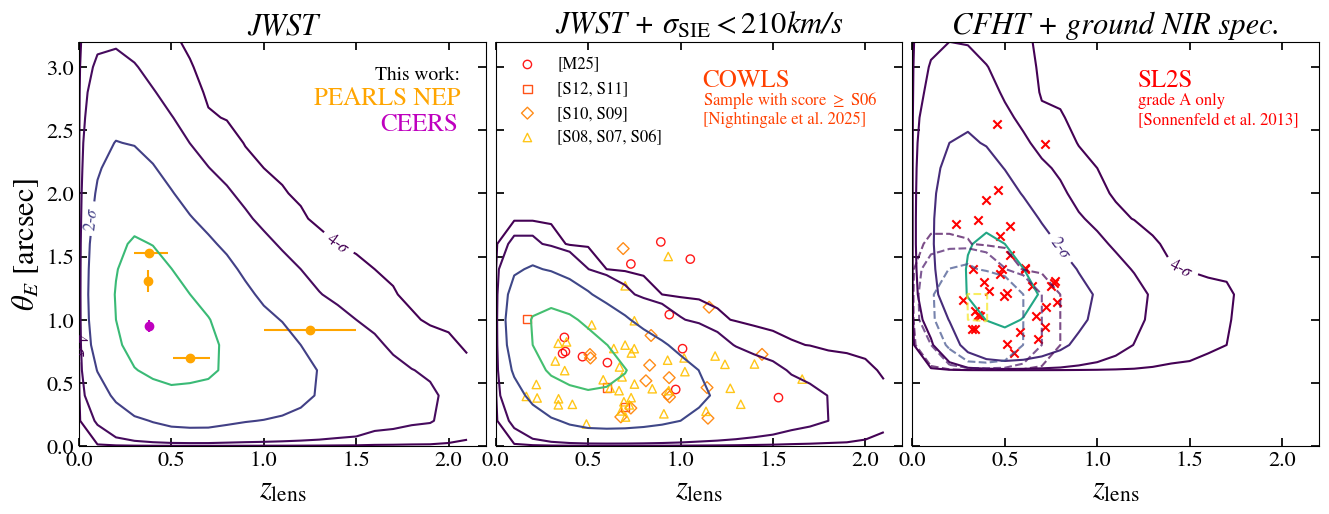}
\caption{Comparison between the Einstein radius vs lens redshift distribution in our sample of lenses (left panel, orange and purple points), in COWLS (central panel, markers are mapped to the confidence score of each lens, S06 being the lowest and M25 the highest) and SL2S (right panel, red crosses).
On top of each panel is reported the telescope used to conduct the relative lens search.
The contours represent the iso-probability levels predicted using the {\tt galess} analytical model (\citealt{Ferrami_lensstat}), including a cut in velocity dispersion ($\sigma_{\rm{SIE}}<210$ km/s) in the central panel.
The shaded dashed contours in the right panel represent the expected distribution in SL2S if we assume the cut in velocity dispersion required to reproduce the distribution of COWLS.}
\label{fig:comp_zl_theta_dist}
\end{figure*}

In Figure \ref{fig:comp_distributions} we compare the distributions in lens and source redshifts, and Einstein radius found in our sample to the Strong Lensing Legacy Survey (SL2S, \citealt{Cabanac_SL2S}), which is based on a algorithmic lens search in the Canada–France–Hawaii Telescope Legacy Survey (CFHTLS) with follow-up ground based spectroscopy, and the COSMOS-Web Lens Survey (COWLS, \citealt{COWLS_I}), which consists in a visual inspection of the COSMOS-Web JWST field.
Figure \ref{fig:comp_zl_theta_dist} shows the link between the lens redshift and Einstein radius distributions for the three surveys introduced before, where the contours are calculated using {\tt galess}\footnote{\url{https://github.com/Ferr013/GALESS}} \citep{Ferrami_lensstat}. 

\begin{figure*}
\includegraphics[height=0.38\linewidth]{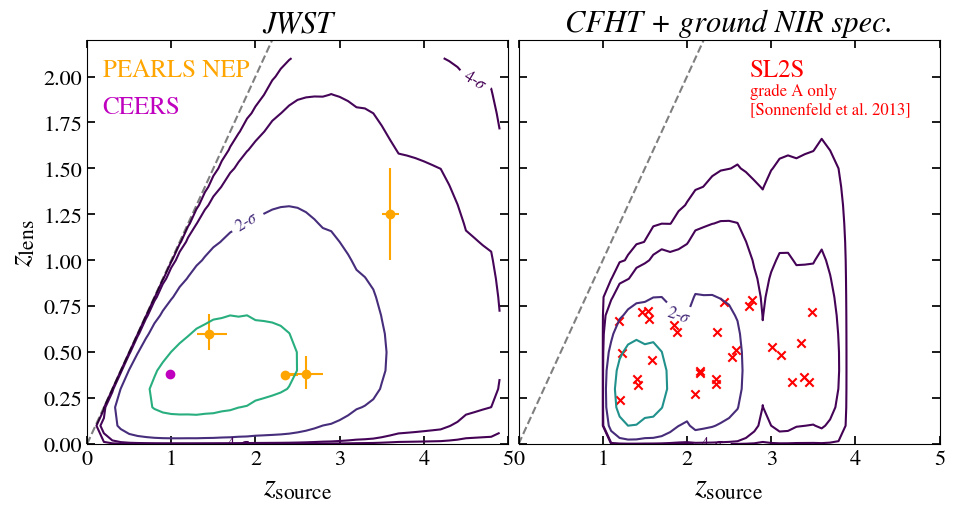}
\caption{Deflector vs source redshift distribution in our sample (left) and SL2S (right), presented as in Figure \ref{fig:comp_zl_theta_dist}.}
\label{fig:comp_zl_zs_dist}
\end{figure*}

This shows that a lens search conducted on JWST photometry can probe higher redshift of the lens and lower Einstein radii sized compared to ground based surveys.
We note that the distributions observed in the COWLS sample might be explained by a imposing an upper bound in velocity dispersion (i.e. total mass) of the lens.
The cut in velocity dispersion was estimated using the $M_\star-\sigma$ relation from \cite{stellarMass_sigma_rel_Cannarozzo}, noticing that the stellar masses in the COWLS lens sample are nearly all below $\log(M_\star/M_\odot) = 11$ and have mean deflector redshift $\approx 0.95$.
If we apply this cut in velocity dispersion to the expectations for the SL2S sample (shaded dashed contours), we are constrained to a parameter space region smaller than the available observations.
\REPLY{While this analysis considers the effect of cosmic variance over the comoving volume occupied by the deflector population of the COWLS sample, we highlight that the forecasting model relies on yet poorly constrained assumptions on the evolution of the Velocity Dispersion Function (VDF). Therefore, the scarcity of large Einstein radii shown in Figure \ref{fig:comp_zl_theta_dist} could also be explained by a rapid evolution with redshift of the VDF normalisation.}
We consider also the correlation between lens and source redshifts in Figure \ref{fig:comp_zl_zs_dist}.
It is clear that JWST will be able to probe much deeper redshifts compared to ground based surveys (also considering the limits in determining emission lines with near-infrared spectroscopy imposed by the earth atmosphere, which are accounted for in the calculation for the probability contours in SL2S).

When compared to other lens searches, our sample yields an extremely high number of secure lenses per unit area.
Accounting for $\sim 30$\% of cosmic variance (\citealt{Cosmic_Variance_TrentiStiavelli}, calculated considering the fact that our total area, obtained combining PEARLS NEP TDF and CEERS, is non contiguous), the number of high-confidence candidates in our sample leads to a density of \REPLY{$125\pm37$} galaxy scale lenses per square degree, equivalent to a strong lens candidate every three to \REPLY{four} NIRCam pointings (9 arcmin$^2$ each). These number are lower bounds depending on the completeness of our search \REPLY{and on the accuracy of the photometric redshift measurements (that ruled out potential candidates as CEERS J141855.6+524527)}.
These densities are compatible with the theoretical predictions of \cite{Holloway_IRsurveys} and \cite{Ferrami_lensstat}.
As a comparison, the SL2S lens search yielded 36 secure lenses over 170 deg$^2$, while the search conducted in DES by \cite{Jacobs_CNN_DES} yielded 500 candidates over 5000 deg$^2$ (of which 138 are spectroscopically confirmed, \citealt{AGEL_DR2}).
The recent lens search conducted in the COSMOS-Web JWST field yielded 17 secure lenses over an area of $0.54$ deg$^2$ \citep{COWLS_I}, with about 400 more with varying degrees of confidence. 
Our results suggest that there should be \REPLY{approximately a} hundred true positive amongst the COWLS candidates, as discussed in \cite{COWLS_III}.

\begin{table*}
  \caption{Catalogue of the lens candidates presented in this work.}
  \label{tab::Discovered_lenses}
  \centering
  \renewcommand{\arraystretch}{1.4}
  \begin{tabular*}{\linewidth}{@{\extracolsep{\fill}} c c c c c }
      \hline 
      Name & R.A. & DEC & $z_{\rm{lens}}^{\text{phot}}$ & $z_{\rm{source}}^{\text{phot}}$\\
      (1) & (2) & (3) & (4) & (5)\\
      \hline 
      \hline 
      PEARLS J172311.9+655014  & 17:23:11.93 & +65:50:14.32 & $^\dagger0.3741_{-0.0002}^{+0.0002}$ & $2.35_{-0.15}^{+0.05}$\\
      PEARLS J172238.9+655143  & 17:22:38.98 & +65:51:43.16 & $1.25_{-0.25}^{+0.25}$ & $3.60_{-0.10}^{+0.10}$\\
      PEARLS J172243.9+655622  & 17:22:43.90 & +65:56:22.97 & $0.60_{-0.11}^{+0.09}$ & $1.45_{-0.21}^{+0.14}$\\
      \REPLY{PEARLS J172339.6+654936}  & 17:23:39.60 & +65:49:35.75 & $0.38_{-0.10}^{+0.08}$ & $2.60_{-0.20}^{+0.20}$\\
      \hline
      CEERS J142031.8+525822  & 14:20:31.80 & +52:58:22.90 & $^\ddagger0.380_{-0.005}^{+0.005}$ & $0.99_{-0.05}^{+0.05}$\\
      \hline
      \hline
  \end{tabular*}
  \begin{flushleft}
  \footnotesize \textit{Notes.} The columns indicate the (1) lens name, (2) right ascension and (3) declination, (4) photometric redshifts of the deflector and (5) of the background source.
  \textit{Comments.} $^\dagger$This deflector has spectroscopic redshift measured with Binospec at the MMT [Willmer, personal communication]. $^\ddagger$This system was also identified as a probable lens (grade B) in SL2S \citep{Sonnenfeld_SL2S_2013}, with the measured spectroscopic redshifts reported in the table. Our estimate of the photo-z for this galaxies has a large error due to NIRCam blue module gap: $z_{\rm{lens}}^{\rm{phot}} = 0.1_{-0.05}^{+0.4}$, $z_{\rm{source}}^{\rm{phot}} = 0.75_{-0.33}^{+0.63}$.
  \end{flushleft} 
\end{table*}

\begin{table*}
  \caption{Lens model parameters and stellar mass for the 4 best lens candidates.}
  \label{tab::Lens_models_parameters}
  \centering
  \renewcommand{\arraystretch}{1.4}
  \begin{tabular*}{\linewidth}{@{\extracolsep{\fill}} c c c c c c c}
      \hline 
      Name & $\theta_E$ ["] & q & P.A. [$^\circ$] & $\gamma_1$ & $\gamma_2$& $\sigma^{\text{SIE}}$ [km s$^{-1}$]\\
      (1) & (2) & (3) & (4) & (5) & (6) & (7)\\
      \hline 
      \hline 
      PEARLS J172311.9+655014 & $1.31_{-0.08}^{+0.09}$ & $0.51_{-0.05}^{+0.02}$ & $-73.9_{-12.4}^{+10.2}$ & $0.02_{-0.01}^{+0.02}$ & $0.05_{-0.02}^{+0.04}$ & $225_{-33}^{+14}$\\
      PEARLS J172238.9+655143 & $0.92_{-0.01}^{+0.02}$ & $0.38_{-0.01}^{+0.01}$ & $46.4_{-1.9}^{+0.8}$ & $0.04_{-0.01}^{+0.02}$ & $0.10_{-0.01}^{+0.01}$ & $247_{-5}^{+6}$\\
      PEARLS J172243.9+655622 & $0.70_{-0.01}^{+0.01}$ & $0.79_{-0.02}^{+0.02}$ & $42.8_{-4.2}^{+1.9}$ & $-0.07_{-0.01}^{+0.01}$ & $0.23_{-0.01}^{+0.01}$ & $204_{-20}^{+35}$\\
      \REPLY{PEARLS J172339.6+654936} & $1.525_{-0.006}^{+0.005}$& $0.62_{-0.03}^{+0.02}$ & $-43.8_{-1.2}^{+3.7}$& $0.02_{-0.01}^{+0.01}$ & $-0.05_{-0.01}^{+0.01}$ & $208_{-13}^{+12}$\\
      \hline
      CEERS J142031.8+525822 & $0.95_{-0.01}^{+0.01}$ & $0.53_{-0.01}^{+0.01}$ & $38.4_{-0.4}^{+0.3}$ & $0.29_{-0.01}^{+0.01}$ & $0.15_{-0.01}^{+0.01}$ & $223_{-17}^{+9}$\\
      \hline
      \hline
  \end{tabular*}
  \begin{flushleft}
  \footnotesize \textit{Notes.} The columns indicate the (1) lens name; (2) Einstein radius in arcseconds, (3) lens axis ratio and (4) position angle; the (5)(6) external shear components; and (7) deflector velocity dispersion entering the isothermal mass profile, in km/s.   
  \end{flushleft} 
\end{table*}

\begin{table*}
  \caption{Total and stellar mass within the Einstein radius.}
  \label{tab::Mass_within_REinstein}
  \centering
  \renewcommand{\arraystretch}{1.4}
  \begin{tabular*}{\linewidth}{@{\extracolsep{\fill}} c c c c c}
      \hline 
      Name & $\theta_E$ [kpc] & 
      $\log M_{\rm{tot}, \leq \theta_E}/M_\odot$ & 
      $\log M_{\star, \leq \theta_E}^{\rm{Kroupa}}/M_\odot$ &
      $\left(M_{\star}/M_{\rm{tot}}\right)_{\leq \theta_E}$\\
      (1) & (2) & (3) & (4) & (5)\\
      \hline 
      \hline 
        PEARLS J172311.9+655014 & $7.27_{-0.41}^{+0.01}$ & $11.54_{-0.05}^{+0.06}$ & $11.31_{-0.17}^{+0.17}$ & $0.58_{-0.24}^{+0.39}$ \\
        PEARLS J172238.9+655143 & $7.67_{-0.38}^{+0.01}$ & $11.61_{-0.11}^{+0.10}$ & $11.11_{-0.08}^{+0.07}$ & $0.31_{-0.11}^{+0.16}$ \\
        PEARLS J172243.9+655622 & $4.68_{-0.51}^{+0.01}$ & $11.23_{-0.14}^{+0.17}$ & $10.59_{-0.12}^{+0.12}$ & $0.23_{-0.11}^{+0.19}$ \\
        \REPLY{PEARLS J172339.6+654936} & $7.94_{-1.49}^{+0.01}$ & $11.61_{-0.13}^{+0.09}$ & $11.37_{-0.15}^{+0.18}$ & $0.57_{-0.24}^{+0.60}$\\
      \hline
        CEERS J142031.8+525822 & $4.95_{-0.34}^{+0.01}$ & $11.34_{-0.07}^{+0.06}$ & $10.45_{-0.20}^{+0.12}$ & $0.13_{-0.06}^{+0.07}$  \\
      \hline
      \hline
  \end{tabular*}
  \begin{flushleft}
  \footnotesize \textit{Notes.} The columns indicate the (1) lens name; the (2) Einstein radius in kpc; the (3) total projected mass and the (4) projected stellar mass within the Einstein radius; and (5) the stellar-to-total mass ratio.   
  \end{flushleft} 
\end{table*}

\section{Discussion and conclusions}\label{Section:Conclusions}

We performed a search for galaxy-scale strong lenses in 144 arcmin$^2$ of JWST NIRCam data, combining the PEARLS NEP TDF and CEERS fields, each containing observations in 9 photometric filters down to mag 28.5.
We first visually inspected each NIRCam pointing, isolating 16 lens candidates, which we subsequentially tested for having different redshift between deflector and source with SED modelling. 
\REPLY{Using a annular mask on each cutout, we fitted the lens and source light with S\'ersic profiles and attempted to reproduced the observed image configurations with a SIE mass distribution plus external shear. 
To test whether the central portion of the mask and the light profile had any impact on our recovered sample, we remodelled the sample using a Multi-Gaussian Expansion to model the lens and source light down to the centre of the deflector. This technique lead to a lensing model of two additional systems compared to the previous step, performing particularly well on lenses with small angular size or faint lensed images compared to the lens light. 
Combining the requirements of a lens model and photo-zs compatible with lensing geometry, we obtained} a sample of \REPLY{5 high-confidence} lenses, corresponding to a density of \REPLY{$125\pm37$} lenses per deg$^2$.
We study the redshift and angular size distributions of this sample, comparing it with past searches and theoretical expectations. 
We confirm that a large fraction of the lenses discoverable by JWST ($\sim$ 30\%) will be above redshift $z_{\rm{lens}}>1$ and have magnified sources with $z_{\rm{source}}>3$, such as PEARLS J172238.9+655143 or the COSMOS-Web Ring (\citealt{MercierCosmosRing, vanDokkumCosmosRing,  ShuntovCosmosRing}), and most Einstein radii smaller than 1".
Given the results of this article, we expect $\sim \REPLY{30}$ of these high-z lenses in COWLS (\citealt{COWLS_I}) or a single Pure Parallel program (equivalent to around 200 NIRCam pointings).
Such a set of high redshift lenses can be used to constrain the evolution of the elliptical galaxies mass function at redshift range higher what is available now.
A sample of lensing systems with high redshifts and small angular sizes can also be used to train neural network lens finders on a range of parameters that has not been explored so far.

%%%%%%%%%%%%%%%%%%%%%%%%%%%%%%%%%%%%%%%%%%%%%%%%%%
\section*{acknowledgments}
\REPLY{We thank the referee, James W. Nightingale, for useful comments that improved the presentation.}
G.F and J.S.B.W. acknowledge the support of the Australian Research Council Centre of Excellence for All Sky Astrophysics in 3 Dimensions (ASTRO 3D), through project number CE170100013, and the
We acknowledge support from the ERC Advanced Investigator Grant EPOCHS (788113), as well as two studentships from STFC.
J.M.D. acknowledges support from project PID2022-138896NB-C51 (MCIU/AEI/MINECO/FEDER, UE) Ministerio de Ciencia, Investigación y Universidades.
This work is based on observations made with the NASA/ESA Hubble Space Telescope (HST) and NASA/ESA/CSA James Webb Space Telescope
(JWST) obtained from the Mikulski Archive for Space Telescopes (MAST) at the Space Telescope Science Institute (STScI), which is operated by the Association of Universities for Research in Astronomy, Inc., under NASA contract NAS 5-03127 for JWST, and NAS 5–26555 for HST. The observations used in this work are associated with JWST programme 2738 and raw images are available on the MAST archive. The authors thank all involved in the construction and operations of the telescope as well as those who designed and executed these observations, their number are too large to list here and without each of their continued efforts, such work would not be possible.

\section*{Software}
This work used the following software packages:
{\tt astropy} \citep{astropy},  
{\tt pypher} \citep{boucaud2016},
{\tt galfit} \citep{Peng2010},
{\tt EAZY} \citep{Brammer2008},
{\tt BAGPIPES} \citep{Carnall2018},
{\tt lenstronomy} \citep{Birrer2018,Birrer2021},
\REPLY{{\tt pyautolens} \citep{Pyautolens_I, Pyautolens_II, Pyautolens_III}},
{\tt galess} \citep{Ferrami_lensstat},
{\tt NumPy} \citep{harris2020array},
{\tt Matplotlib} \citep{Hunter2007},
{\tt SciPy} \citep{2020SciPy-NMeth}.

\section*{Data Availability}
JWST data from PEARLS programme ID 2738 is publicly available on the MAST portal and accessible at \href{https://mast.stsci.edu/portal/Mashup/Clients/Mast/Portal.html?searchQuery=%7B%22service%22%3A%22CAOMBYPROPID%22%2C%22inputText%22%3A%222738%22%2C%22paramsService%22%3A%22Mast.Caom.Filtered%22%2C%22title%22%3A%22Proposal%20IDs%3A%202738%22%2C%22columns%22%3A%22*%22%2C%22caomVersion%22%3Anull%7D}{this link}. 
Accompanying HST data from TREASUREHUNT is available from \url{https://doi.org/10.17909/wv13-qc14}.
JWST data from CEERS is publicly available at \url{https://ceers.github.io/dr07.html}.
\REPLY{The FITS cutouts for all wavebands, including RMS noise maps and PSFs for all 16 candidates are made available at \url{https://zenodo.org/records/15846915}.}



\bibliographystyle{mnras}
\bibliography{bib_lens_search} % if your bibtex file is called example.bib

%%%%%%%%%%%%%%%%%%%%%%%%%%%%%%%%%%%%%%%%%%%%%%%%%%

%%%%%%%%%%%%%%%%% APPENDICES %%%%%%%%%%%%%%%%%%%%%

\appendix

\section{Initial sample of lens candidates}
\begin{figure*}
\includegraphics[width=\linewidth]{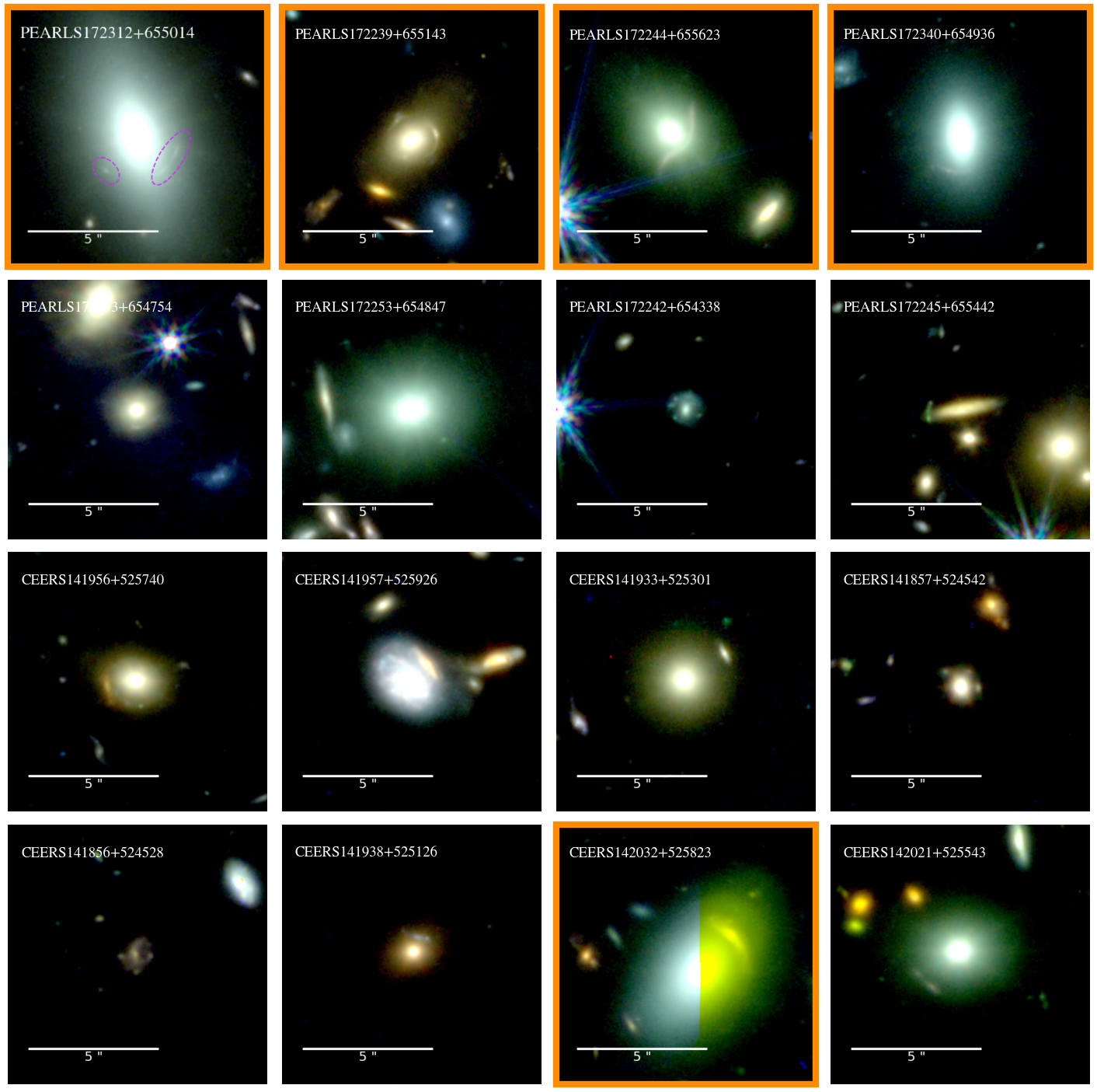}
\caption{Initial sample of lens candidates in {PEARLS} NEP field and CEERS fields, shown as 10"$\times$10" cutouts centred on the candidate deflector. The systems highlighted with an orange border are the \REPLY{5} high-confidence strong lenses with measured photometric redshifts and a lens model.}
\label{fig:lens_candidates_initial}
\end{figure*}

Figure \ref{fig:lens_candidates_initial} contains the RGB composite images of the initial sample of 16 candidate lenses obtained by visual inspection over the PEARLS NEP TDF and CEERS JWST fields. The high confidence candidates, are highlighted with an orange border. Their photometric redshift measurements, where available, are listed in Table \ref{tab:Photo_z_all_candidates}.
Table \ref{tab:Photo_z_all_candidates} also lists the criteria that excluded a candidate from the final list of high probability lenses. If either of the photometric redshift is missing, or the photo-z measurements are incompatible with a lensing geometry, the rejection criteria is flagged as `photo-z', while if the simple lens mass (SIE + external shear) or source light (single Sèrsic profile) models employed are not able to reproduce the observed configuration the rejection criteria is flagged as `lens model'.
\REPLY{After subtracting the lens light with a MGE, the systems CEERS J141856.7+524542, CEERS J141956.2+525740, PEARLS J172233.1+654753 appear as the best candidates for follow-up confirmation out of the 11 non-confirmed lensing systems.}

\begin{table*}
  \caption{Photometric redshift measurements for the 16 systems flagged as candidate lenses in the visual inspection process.}
  \label{tab:Photo_z_all_candidates}
  \centering
  \renewcommand{\arraystretch}{1.4}
  \begin{tabular*}{0.95\linewidth}{@{\extracolsep{\fill}} c c c c c c c}
      \hline 
      Name & R.A. & DEC & $z_{\rm{lens}}^{\text{phot}}$ & $z_{\rm{source}}^{\text{phot}}$ & Rejection & Comments\\
           &      &     &                     &                     & Criteria  &         \\
      \hline 
      \hline 
      PEARLS J172311.9+655014 & 17:23:11.93 & +65:50:14.32 & $0.3741_{-0.002}^{+0.002}$ & $2.35_{-0.15}^{+0.05}$ & - & $^{(a)}$\\
      PEARLS J172238.9+655143 & 17:22:38.98 & +65:51:43.16 & $1.25_{-0.25}^{+0.25}$ & $3.60_{-0.10}^{+0.10}$ & - & $^{(b)}$\\
      PEARLS J172243.9+655622 & 17:22:43.90 & +65:56:22.97 & $0.60_{-0.11}^{+0.09}$ & $1.45_{-0.21}^{+0.14}$ & - & \\
      PEARLS J172339.6+654936 & 17:23:39.60 & +65:49:35.75 & $0.38_{-0.10}^{+0.08}$ & $2.60_{-0.20}^{+0.20}$ & \REPLY{-} & \\ %NEP4
      PEARLS J172233.1+654753 & 17:22:33.16 & +65:47:53.66 & $\left[0.57_{-0.15}^{+0.15}\right]$ & $\left[2.90_{-0.14}^{+0.14}\right]$ & lens model & $^{(c)}$\\ %NEP6
      PEARLS J172253.1+654847 & 17:22:53.14 & +65:48:47.42 & $\left[0.55._{-0.1}^{+0.1}\right]$ & $\left[0.65._{-0.1}^{+0.1}\right]$ & photo-z & $^{(d)}$\\ %NEP7
      PEARLS J172242.6+654338 & 17:22:42.68 & +65:43:38.02 & $\left[0.63._{-0.1}^{+0.1}\right]$ & - & both & $^{(e)}$ \\ %NEP8
      PEARLS J172245.2+655441 & 17:22:45.26 & +65:54:41.96 & $\left[0.98._{-0.1}^{+0.1}\right]$ & - & both & $^{(e)}$ \\ %NEP15
      \hline
      CEERS J141956.2+525740  & 14:19:56.24 & +52:57:40.80 & - & - & both & \\ %CRS1
      CEERS J141957.3+525926  & 14:19:57.30 & +52:59:26.22 & - & - & both & \\ %CRS2
      CEERS J141933.3+525300  & 14:19:33.33 & +52:53:00.90 & $\left[1.13_{-0.17}^{+0.19}\right]$ & $\left[1.51_{-0.13}^{+0.15}\right]$ & lens model & $^{(f)}$\\  %CRS3
      CEERS J141856.7+524542  & 14:18:56.75 & +52:45:42.45 & - & - & photo-z & $^{(g)}$ \\ %CRS6
      CEERS J141855.6+524527  & 14:18:55.60 & +52:45:27.80 & $\left[1.55_{-0.18}^{+0.18}\right]$ & $\left[1.61_{-0.18}^{+0.18}\right]$ & photo-z & $^{(d)}$\\
      CEERS J141937.5+525125  & 14:19:37.53 & +52:51:25.97 & - & - & lens model& \\ %CRS8
      CEERS J142031.8+525822  & 14:20:31.80 & +52:58:22.90 & $0.380_{-0.005}^{+0.005}$ & $0.99_{-0.05}^{+0.05}$ & - & $^{(h)}$\\ %CRS9
      CEERS J142021.4+525543  & 14:20:21.48 & +52:55:43.12 & $\left[0.73_{-0.24}^{+0.03}\right]$ & $\left[4.21_{-0.21}^{+0.14}\right]$ &lens model & \\ %CRS10
      \hline
      \hline
  \end{tabular*}
  \begin{flushleft}
  \footnotesize \textit{Notes.} From left to right, the columns indicate the lens name, right ascension and declination, photometric redshifts of the deflector and of the background source, and the criteria that excluded the system from the high-confidence sample (either photo-z or lens modeling). 
  The systems without both redshift measurements or with measurements listed in square brackets are not included in the final sample of high-confidence lens candidates.
  \REPLY{PEARLS J172311.9+655014 and PEARLS J172238.9+655143 are identified as radio sources in \citealt{HyunTDFRadio} with index 382 and 182, respectively.}\\
  \textit{Comments.} 
  $^{(a)}$This deflector has spectroscopic redshift measured with Binospec at the MMT [Willmer, personal communication]. 
  $^{(b)}$This source also lies in a NIRISS pointing, and the spectral extraction (\citealt{Estrada-Carpenter2024}) yields redshift of $z_{\rm{lens}}=1.258\pm0.005$ (more details will be presented in \citealt{NathanAdams_EinsteinRingInPEARLS}); 
  $^{(c)}$Both deflector and source photo-z are bimodal ($z_{\rm{lens}}$ = 0.57 or 1.82, $z_{\rm{source}}$ = 0.3 or 2.9). Suspected arc may be part of diffraction spike (clover pattern) of galaxy bulge; 
  $^{(d)}$Source just behind foreground galaxy; 
  $^{(e)}$Starburst inside host; 
  $^{(f)}$The SED shows a 1.6 micron bump at z=1.0-1.2 not 1.5, these may be companion/minor merger; 
  $^{(g)}$The three point images are part of diffraction spike; 
  $^{(h)}$Target in the blue module gap. This lens was also found in SL2S with spectroscopic redshifts for deflector and source, reported in the table. Our estimate of the photo-z for this galaxies has a large error due to NIRCam blue module gap: $z_{\rm{lens}}^{\rm{phot}} = 0.1_{-0.05}^{+0.4}$, $z_{\rm{source}}^{\rm{phot}} = 0.75_{-0.33}^{+0.63}$.
  \end{flushleft} 
\end{table*}

%%%%%%%%%%%%%%%%%%%%%%%%%%%%%%%%%%%%%%%%%%%%%%%%%%

% Don't change these lines
\bsp	% typesetting comment
\label{lastpage}
\end{document}